\documentclass{elsart}
\usepackage{graphicx}
\usepackage{listings}
\newcommand{\bq}{\begin{equation}}
\newcommand{\eq}{\end{equation}}
\newcommand{\cma}{~~,}
\newcommand{\eos}{~~.}
\newcommand{\bunt}{\itshape}

\newcommand{\ket}[1]{\big\arrowvert{#1}\bigr\rangle}

\begin{document}
\lstset{labelstep=1,language=C++,labelstyle=\tiny,basicstyle={\ttfamily\footnotesize},keywordstyle={\bfseries},aboveskip=0.2cm,belowskip=0.2cm}
\begin{frontmatter}
\title{Parallelization Strategies for\\
  Density Matrix Renormalization Group\\
  Algorithms on Shared-Memory Systems}

\author[hager]{G.~Hager},
\author[eric]{E.~Jeckelmann},
\author[holger]{H.~Fehske} and
\author[hager]{G.~Wellein}
\address[hager]{Regionales Rechenzentrum Erlangen (RRZE),
    Martensstra{\ss}e 1, D-91058 Erlangen, Germany}
\address[eric]{Johannes-Gutenberg-Universit\"at Mainz, Institut f\"ur 
    Physik (Gruppe KOMET 337), Staudingerweg 7/9, D-55099 Mainz,
    Germany}
\address[holger]{Ernst-Moritz-Arndt-Universit\"at
    Greifswald, Institut f\"ur Physik, Domstr.\ 10a, D-17489
    Greifs\-wald, Ger\-many}

\begin{abstract}
  Shared-memory parallelization (SMP) strategies for density matrix
  renormalization group (DMRG) algorithms enable the treatment of
  complex systems in solid state physics.  We present two different
  approaches by which parallelization of the standard DMRG algorithm
  can be accomplished in an efficient way. The methods are illustrated
  with DMRG calculations of the two-dimensional Hubbard model and the
  one-dimensional Holstein-Hubbard model on contemporary SMP
  architectures. The parallelized code shows good scalability up to at
  least eight processors and allows us to solve problems which exceed
  the capability of sequential DMRG calculations.
\end{abstract}

\end{frontmatter}

\section{Introduction}

During the past decade the density matrix renormalization group (DRMG)
\cite{whi92,whi93} has been established as a powerful numerical
technique for solving many-body problems in Interacting Electron
Systems, Classical Statistical Physics, and Quantum Chemistry (for a
review, see~\cite{pesch99})\@.  For strongly correlated quantum lattice
systems, DMRG techniques complement and sometimes even replace
traditional methods like exact diagonalization (ED) or (Quantum) Monte
Carlo algorithms \cite{jec98}\@.
In particular, for quantum systems in one spatial dimension and
short-range interactions DMRG provides approximations to the ground
state, to the low-lying excited states and to spectral properties with
high accuracy at a modest computational effort.

Although the computational requirements can easily exceed the
resources of sequential computers such as PCs or workstations and grow
rapidly with increasing complexity (dimensionality or range of
interactions), no efficient parallelization approach has been
introduced for DMRG algorithms to date.
Starting from a sequential C++ package developed by White and
Jeckelmann we discuss two shared-memory parallelization strategies for
the superblock diagonalization and analyze their scalability and
performance on state of the art SMP systems like IBM p690, SGI
Origin and Intel Itanium2-based servers.

The first approach uses the inherent parallelism contained in the
dominating operation of all DMRG calculations, i.e.\ dense
matrix-matrix multiplication (generally implemented 
as a call to the BLAS subroutine DGEMM)\@. This is the lowest
possible parallelization level and is thus prone to scale badly,
especially in certain numerical limits where the matrices tend to be
small and the parallelization overhead becomes dominant. It is
nevertheless a viable strategy when the numerical structure of the
problem at hand is appropriate. In those rare cases, a significant
percentage of theoretical peak performance can be achieved.

The second approach targets the parallelization of the sparse
matrix-vector multiplication contained in the superblock
diagonalization algorithm and overcomes the overhead-induced
performance bottlenecks of parallel DGEMM\@. Here we use OpenMP to
achieve better scalability at the cost of an only slightly more
complicated code structure. DMRG calculations can then be carried out
with sufficient parallel efficiency on up to eight CPUs,
depending on the physical problem. Modern supercomputer architectures of
the SMP cluster type with large-memory SMP nodes comprising up
to eight or sixteen CPUs are the primary target systems on which this approach
can be used with success.

In the following section we will very briefly discuss the basics of
the DMRG algorithm and its implementation in the software package used
as a starting point for parallelization. Section \ref{par} deals with
the two different parallelization strategies and gives some estimates
about expected performance. In section \ref{per} we will present
the results of performance measurements on different architectures.
Section \ref{phys} then illustrates the benefits of parallel DMRG,
introducing some physical results not obtainable with ED methods.
Finally, section \ref{conc} gives some conclusions and summarizes what
has yet to be done.

\section{DMRG Algorithm}\label{alg}

\subsection{Basics}

Determining the ground state and spectral properties of interacting quantum
lattice models like e.~g.\ the Hubbard model
\begin{equation}\label{hm}
H_{\mathrm{HM}} = 
-t\sum_{\left<ij\right>,\sigma}\left[c^\dag_{i\sigma}c_{j\sigma}+
  \mathrm{H.c.}\right]+U\sum_in_{i\uparrow}n_{i\downarrow}\cma
\end{equation}
or the Holstein-Hubbard model
\begin{equation}
\label{hhm}
H_{\mathrm{HHM}} =  
H_{\mathrm{HM}} +g\omega_0\sum_{i,\sigma}(b_i^\dag+b_i)n_{i\sigma}+
\omega_0\sum_ib_i^\dag b_i 
\end{equation}
is a fundamental but difficult problem in solid-state physics.  Here,
$c^\dag_{i\sigma}$ ($c_{i\sigma}$) denote fermionic creation
(annihilation) operators of spin-$\sigma$ ($\sigma\in\{\uparrow,\downarrow\}$)
electrons, $n_{i\sigma}=c^\dag_{i\sigma}c_{i\sigma}$, and
$b_i^\dag$ ($b_i$) are the corresponding bosonic phonon creation
(destruction) operators (for the construction of the Hilbert space
basis see, e.~g., Ref.~\cite{WRF96}).

The Hubbard model, independently proposed in 1963 by Gutzwiller,
Hubbard and Kanamori~\cite{GHK63}, was originally designed to describe the
ferromagnetism of transition metals. The physics of the model is
governed by the competition between electron itinerancy ($t$;
delocalization) and short--range Coulomb repulsion ($U$; localization,
magnetic order), where the fermionic nature of the charge carriers is
of great importance (Pauli exclusion principle, i.e., the existence of
an `effective' long--range interaction). Besides the ratio $U/t$, the
particle density $n$, the temperature $T$, and the spatial dimension
$D$ (geometry of the lattice) are crucial parameters involved in the
model. Successively, the Hubbard model was studied in the context of
magnetism, metal--insulator (Mott) transition, heavy fermions and
high-temperature superconductivity as the probably most simple model
to account for strong correlation effects.

In addition to the purely electronic interactions in the Hubbard model
it is often necessary to incorporate the coupling to lattice degrees
of freedom to describe the electronic properties of solids. In the
Holstein-Hubbard model~\cite{Ho59}, the second term couples 
the electronic system
locally to an internal optical degree of freedom of the effective
lattice, whereas the third term takes into account the elastic energy
of a harmonic lattice. $g$ and $\omega_0$ denote the electron-phonon
coupling constant and the phonon frequency, respectively.  In the
single-electron case, the Holstein model has been studied extensively
as a paradigmatic model for polaron formation.  At half-filling the
electron-phonon coupling may lead to a Peierls instability (in
competition to the antiferromagnetic instability triggered by $U$).

Although a tremendous amount of work has been devoted to the solution
of the Hubbard and Holstein-Hubbard models, exact results are very
rare and only a few special cases and limits have so far been
understood analytically.  Therefore a numerical treatment of both
models seems to be inevitable.

Due to the locality of interactions, the matrix representation of the
Hamiltonian operator $H$ in a real-space basis is generally very
sparse. In an ED approach this matrix is (partially) diagonalized with
Lanczos, Davidson or similar algorithms. The dominant operation is
then a sparse matrix-vector multiplication (MVM) of $H$ with some
vector $\vec v$\@. Due to the exponential growth of degrees of freedom
with increasing system size, ED methods are limited to relatively
small systems and generally require vast computing resources and
memory bandwidth.

The DMRG algorithm \cite{whi92,whi93,nw99} tries to overcome those
drawbacks by implementing a variational scheme that truncates the
Hilbert space used to represent $H$ in an optimal way. It is the
selection of the basis states that lays the groundwork on which DMRG
is built.

\subsection{The Algorithm}

\begin{figure}
  \centerline{\setlength{\unitlength}{4144sp}%
\begingroup\makeatletter\ifx\SetFigFont\undefined%
\gdef\SetFigFont#1#2#3#4#5{%
  \reset@font\fontsize{#1}{#2pt}%
  \fontfamily{#3}\fontseries{#4}\fontshape{#5}%
  \selectfont}%
\fi\endgroup%
\begin{picture}(3416,793)(3650,-3297)
\thinlines
\special{ps: gsave 0 0 0 setrgbcolor}\put(3741,-3205){\framebox(1278,319){}}
\special{ps: gsave 0 0 0 setrgbcolor}\put(5059,-3205){\framebox(1916,319){}}
\special{ps: gsave 0 0 0 setrgbcolor}\put(5538,-2648){\vector(-2,-1){286.400}}
\special{ps: grestore}\special{ps: gsave 0 0 0 setrgbcolor}\put(3722,-3225){\oval(120,120)[bl]}
\put(3722,-2866){\oval(120,120)[tl]}
\put(6994,-3225){\oval(120,120)[br]}
\put(6994,-2866){\oval(120,120)[tr]}
\multiput(3722,-3285)(114.80702,0.00000){29}{\line( 1, 0){ 57.404}}
\multiput(3722,-2806)(114.80702,0.00000){29}{\line( 1, 0){ 57.404}}
\multiput(3662,-3225)(0.00000,102.57143){4}{\line( 0, 1){ 51.286}}
\multiput(7054,-3225)(0.00000,102.57143){4}{\line( 0, 1){ 51.286}}
\put(4100,-3125){\makebox(0,0)[lb]{\smash{\SetFigFont{14}{16.8}{\familydefault}{\mddefault}{\updefault}\special{ps: gsave 0 0 0 setrgbcolor}system\special{ps: grestore}}}}
\put(5498,-3125){\makebox(0,0)[lb]{\smash{\SetFigFont{14}{16.8}{\familydefault}{\mddefault}{\updefault}\special{ps: gsave 0 0 0 setrgbcolor}environment\special{ps: grestore}}}}
\put(5458,-2648){\makebox(0,0)[lb]{\smash{\SetFigFont{14}{16.8}{\familydefault}{\mddefault}{\updefault}\special{ps: gsave 0 0 0 setrgbcolor}superblock\special{ps: grestore}}}}
\end{picture}}
\caption{\label{blocks}\bunt Division of the complete physical system into
  ``system block" and ``environment block".  Both blocks together form
  the ``superblock" whose Hamiltonian matrix is diagonalized.}
\end{figure}
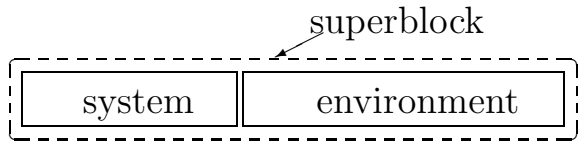

DMRG splits the physical system (usually in real space, although
a momentum space approach is possible) into two pieces, the
so-called \emph{system block} and the \emph{environment block}\@.
Both together form the \emph{superblock} (see Fig.~\ref{blocks})\@.

\begin{figure}
  \centerline{\setlength{\unitlength}{4144sp}%
\begingroup\makeatletter\ifx\SetFigFont\undefined%
\gdef\SetFigFont#1#2#3#4#5{%
  \reset@font\fontsize{#1}{#2pt}%
  \fontfamily{#3}\fontseries{#4}\fontshape{#5}%
  \selectfont}%
\fi\endgroup%
\begin{picture}(4605,384)(5803,-2773)
\thinlines
\put(7517,-2597){\circle*{206}}
\put(8060,-2600){\circle*{206}}
\special{ps: gsave 0 0 0 setrgbcolor}\put(7124,-2597){\line( 1, 0){1309}}
\special{ps: grestore}\special{ps: gsave 0 0 0 setrgbcolor}\put(5815,-2761){\framebox(1309,360){}}
\special{ps: gsave 0 0 0 setrgbcolor}\put(8433,-2761){\framebox(1963,360){}}
\special{ps: gsave 0 0 0 setrgbcolor}\put(7126,-2761){\dashbox{57}(675,360){}}
\put(9054,-2630){\makebox(0,0)[lb]{\smash{\SetFigFont{14}{16.8}{\familydefault}{\mddefault}{\updefault}\special{ps: gsave 0 0 0 setrgbcolor}$\bar H^\mathrm{R}_{l'-1}$\special{ps: grestore}}}}
\put(6142,-2630){\makebox(0,0)[lb]{\smash{\SetFigFont{14}{16.8}{\familydefault}{\mddefault}{\updefault}\special{ps: gsave 0 0 0 setrgbcolor}$\bar H_{l+1}$\special{ps: grestore}}}}
\end{picture}}
\caption{\label{step}\bunt One step of the finite system DMRG 
algorithm (left-to-right phase)\@. $\bar H_{l+1}$ and
$\bar H_{l'-1}^\mathrm{R}$ are system block and environment block
Hamiltonians in the reduced density matrix eigenbasis.}
\end{figure}
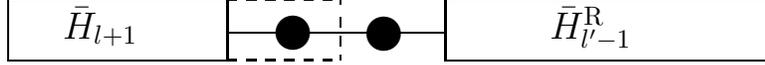

The central entity in the algorithm is the \emph{reduced density
matrix}
\bq\label{rdmat}
\rho_{ii'} = \sum_j \psi^*_{ij}\psi_{i'j}\cma
\eq
where $i$ and $j$ label the states of the system and environment
blocks, respectively, so that a superblock state $\ket\psi$ can be
composed:
\bq
\ket{\psi}=\sum_{ij}\psi_{ij}\ket{i\,}\ket{j\,}\eos
\eq
Definition (\ref{rdmat}) shows that in $\rho$ the states of the
environment block are summed over. In this manner all
possible boundary conditions that the environment may impose
on the system are incorporated in the density matrix. It can now
be shown \cite{nw99} that the eigenstates of $\rho$ with the largest 
eigenvalues are those that have the most significant impact
on observables, i.~e.\ in order to get a good guess at an optimal
basis set for the superblock Hamiltonian one has to 
\begin{itemize}
\item diagonalize the reduced density matrix for a system block of
  size $l$ and extract the $m$ eigenvectors with largest eigenvalue,
\item construct all relevant operators (system block and environment
  Hamiltonians, observables) for a system block of size $l+1$ in the
  reduced density matrix eigenbasis,
\item form a superblock Hamiltonian from the system and environment
  block (size $l-1$) Hamiltonians plus two single sites (see
  Fig.~\ref{step}) and determine its ground state by diagonalization.
\end{itemize}
These steps must be repeated several times, shifting the interface
between system block and environment block back and forth until
some convergence criterion is fulfilled. This might be e.~g.\ 
stationarity of the ground state energy or a sufficiently small
\emph{discarded weight}, which is the sum of all density matrix
eigenvalues that were not considered when forming the basis.
The procedure can be generalized to two dimensions, although
it is not quite clear as to how the best ``path'' for the
sweeps through the grid should be chosen \cite{nw99}\@.

The accuracy of observables like the ground state energy depends on
the number $m$ of density matrix states kept. The discarded weight
gives some hint for choosing the right $m$ for a particular problem.
Usually one starts with $m$ rather small and increases $m$ every time
the ground state energy has converged. Nevertheless most of the
computing time is spent in the sweeps with largest $m$\@. Sensible
values for $m$ depend on the physical model under consideration. In
the one-dimensional case where DMRG usually performs best, $m=500$ to
1000 is often sufficient to get decent data, even for models with
electron-phonon interaction like the HHM (\ref{hhm})\@. In two
dimensions a larger $m$ is in order, e.~g.\ $m=2000$ to 10000
for a 2D Hubbard model (\ref{hm})\@. Although in that case performance 
and memory requirements easily exceed the resources of standard PCs, they 
are still far below those needed for an ED
approach, and valuable results can often be obtained on off-the-shelf
hardware instead of teraflop-class supercomputers (see
section \ref{per})\@.

It must be stressed that many complications show up in implementing
the algorithm for a real-world problem. Fermionic and
bosonic commutation rules, reflection and other symmetries, boundary
conditions, degeneracies etc. all require special attention \cite{jec98,whi93}\@. 
Here we wish to concentrate on the performance and parallelization
aspects alone.

Diagonalization of the superblock Hamiltonian is the most
time-consuming part of the algorithm and is usually done by a Lanczos
or Davidson procedure. Thus repeated multiplications of $H$ with superblock
vectors $\psi$ have to be performed. This is not done by
constructing $H$ explicitly as a matrix, but by using the fact
that a Hamiltonian that describes the concatenation of two blocks
can be written as
\bq
H_{ij;i'j'} = \sum_\alpha A_{ii'}^\alpha B_{jj'}^\alpha\cma
\eq
where $A$ and $B$ are operators in the two blocks and $\alpha$ counts
different terms in the Hamiltonian. Due to the fact that $H$
``lives'' in two blocks and thus has double indices, the MVM
is actually of the matrix-matrix type at the lowest 
level:
\bq\label{hamphi}
\sum_{i'j'}H_{ij;i'j'}\psi_{i'j'} = 
\sum_\alpha\sum_{i'}A^\alpha_{ii'}\sum_{j'}
B^\alpha_{jj'}\psi_{i'j'}\eos
\eq
Dense matrix-matrix multiplication can be optimized using standard
unrolling and blocking techniques \cite{gh01} so that peak performance
is theoretically achievable on modern cache-based RISC architectures.
This is not quite true for very small matrices, where loop overhead
and pipeline fill-up effects come into play, but the MVM part of DMRG is
nevertheless well suited for RISC machines.

A slight complication arises because it is quite unfavourable with
respect to performance and memory requirements to use dense matrices
throughout. Many operators only have nonzero matrix elements between
states with specific quantum numbers (or quantum number differences),
so that it is sufficient to store the nonzero blocks. Those blocks are
labeled by indices $R(k)$ on the RHS and are, by virtue of the MVM,
mapped to blocks with indices $L(k)$ on the LHS\@. Consequently, there
is an additional sum over quantum numbers in (\ref{hamphi})\@.
Omitting the ``normal'' matrix indices, (\ref{hamphi}) becomes
\begin{equation}
  H\psi =  \sum_\alpha\sum_k\left(H\psi\right)^\alpha_{L(k)}
 =  \sum_\alpha\sum_k A_k^\alpha\psi_{R(k)}
  \left[B^\mathrm{T}\right]_k^\alpha\label{hamphib}\eos
\end{equation}
In the software package developed by White and Jeckelmann,
the structure of MVM in the Davidson
algorithm is exactly as shown above, featuring two nested
loops that handle Hamiltonian terms and quantum numbers
separately.

Every shared-memory parallelization attempt must identify loops in the
algorithm that lend themselves to parallel execution.  In
(\ref{hamphib}) three such loops are visible: the innermost
matrix-matrix multiplication (twice), the sum over quantum numbers and
the sum over terms in the Hamiltonian.

\section{Parallelization of the Superblock Diagonalization}\label{par}

As shown in the previous section, the performance of (non-dynamical)
DMRG calculations is governed by the superblock diagonalization
algorithm, in which a sparse MVM plays
the dominant role. Fortunately the basic operation in this sparse MVM
is dense matrix-matrix multiplication, which is well optimized in the form
of BLAS DGEMM on most architectures. Single-CPU performance of DMRG
calculations can potentially achieve a significant fraction of peak
speed.  

SMP parallelization can be performed in a variety of ways, two
of which are targted here: DGEMM threading and OpenMP in the sparse
MVM procedure.

\subsection{Shared-Memory DGEMM parallelization}

This approach is the simplest one possible due to the fact that no
additional programming effort is necessary. Parallel BLAS libraries
exist for virtually all contemporary SMP architectures, thus relinking
with another library is all that is required. All parallelization
complexities are hidden inside vendor-provided DGEMM code.

Unfortunately, the DMRG method has an important drawback ---
the matrices which form the operands for DGEMM calls are
often quite small, leading to non-negligible parallelization
overhead (load imbalance, barrier wait, thread wakeup)\@.
This fact makes the DGEMM approach unsuitable for a large 
class of problems. See section \ref{per} for performance
results.

\subsection{OpenMP Parallelization of MVM}\label{omppar}

One of the basic rules of OpenMP parallelization is to try to find
loops that are as far as possible at the outside of a loop nest and
identify their parallelism.  The sparse MVM at the core of the
Davidson diagonalization routine is a viable target for this approach.

In a first attempt one would simply use an \verb.omp parallel for.
directive at the outer loop of (\ref{hamphib})\@. This, however, yields
unsatisfactory performance because the outer loop goes over the
terms in the Hamiltonian, and although the number of terms can easily
become a couple of hundreds (especially when using a large number
of sites), load imbalance will readily show up. Moreover the number
of terms can become very small in the course of the calculation
when the system block comprises a couple of sites only.

The inner loop over the quantum numbers suffers essentially from the
same deficiencies when it comes to parallelization. In order to get
proper scaling, the loop nest has to be eliminated, leading to a
single loop. This is the original code of the loop
nest:\footnote{The pseudocode snippets in this section are simplified
  excerpts that serve to illustrate the coding strategy. They do not
  constitute runnable code.}
\begin{lstlisting}{}
// W is wave vector, R ist result
for(i=0; i < number_of_hamiltonian_terms; i++)
   {
      term = hamiltonian_terms[i];
      for(q=0; q < term.number_of_blocks; q++)
         {
            li = term[q].left_index;
            ri = term[q].right_index;

            temp_matrix = term[q].B.transpose() * W[ri];
            R[li] += term[q].A * temp_matrix;
         }
   }
\end{lstlisting}
The outer loop is for the Hamiltonian terms whereas the inner loop
counts quantum numbers. The StateSet indices \verb.li. and \verb.ri.
identify blocks with certain quantum numbers in the wave vectors. 
There are some peculiarities one must take care of:
\begin{itemize}
\item Every loop iteration writes to some part of the result vector,
  identified by \verb.li.\@. Parallelization must account for the
  possibility that any two iterations might have the same value for
  \verb.li.\@.
\item The trip count for the inner loop is not a constant but depends
  on the term.
\end{itemize}
So when replacing the loop nest by a single loop, one has to
take some measures with respect to bookkeeping. First,
a prologue loop must prepare an array that stores 
references to all blocks required:
\begin{lstlisting}{}
for (ics=0,i=0; i < number_of_hamiltonian_terms; i++)
   {
     term = hamiltonian_terms[i];
     for(q=0; q < term.number_of_blocks; q++)
        {
           block_array[ics] = &term[q];
           ics++;
        }
   }
icsmax = ics;
\end{lstlisting}
Second, an array of OpenMP locks has to be set up (once) that later
protect from race conditions when updating the result vector.  This
array could potentially be established using a C++ vector class
(dynamic resizeability), but experience shows that most compilers have
severe difficulties in parallelizing OpenMP loops that handle
complicated C++ objects. Thus the necessary arrays were declared as
having a fixed length, and appropriate checking mechanisms (not shown
here) prevent boundary violation:
\begin{lstlisting}{}
static int flag=0;
if(!flag)
   {
      flag=1;
      for(i=0; i < MAX_NUMBER_OF_THREADS; i++)
         mm[i] = new Matrix // temp. matrix
      for(i=0; i < MAX_NUMBER_OF LOCKS; i++)
         {
            locks[i] = new omp_lock_t;
            omp_init_lock(locks[i]);
         }
   }
\end{lstlisting}
Now the loop nest can be transformed into a single parallel loop.
The required temporary matrix for each thread is provided inside 
the parallel region but before the loop actually starts:
\begin{lstlisting}{}
#pragma omp parallel private(mytmat,li,ri,myid,ics)
   {
      myid = omp_get_thread_num();
      mytmat = mm[myid]; // temporary matrix, thread-local
#pragma omp for
      for(ics=0; ics < icsmax; ics++) 
      {
         li = block_array[ics]->left_index; // StateSet indices 
         ri = block_array[ics]->right_index;
    
         
         mytmat = block_array[ics]->B.transpose() * W[ri];

         omp_set_lock(locks[li]);
         R[li] += block_array[ics]->A * mytmat;
         omp_unset_lock(locks[li]);
      }
\end{lstlisting}
Only the second matrix-matrix multiplication has to be protected via OpenMP
locks, as it writes to block number \verb.li. of the result vector.
The first one stores its result in a thread-local temporary matrix.

In our default benchmark case (see following section), sparse MVM
takes about 85\,\% of total computing time in the serial case. We
therefore expect parallel speedups of up to 6 or 7, not taking into
account mutual locking overhead, thread startup and the like.

\section{Performance Results on Contemporary SMP Systems}\label{per}

Two benchmark cases have been investigated in order to
show the performance of the different parallelization 
strategies:
\begin{enumerate}
\item\label{case1} The `default' benchmark case used here, unless
  otherwise noted, is a calculation of ground state properties for the
  Hubbard Model (\ref{hm}) in two dimensions with 4x4 sites and
  periodic boundary conditions (BCs) at half-filling with $U=4$ and
  isotropic delocalization $t_{x,y}=1$\@.  
  Although we stick to $m=2000$ for practical reasons, the number of density
  matrix states kept, $m$, must be larger ($m \approx 7000$) to obtain 
  a good approximation of the ground state wavefunction, in particular, 
  to preserve translational invariance. 
\item\label{case2} The second benchmark, an 8-site one-dimensional 
  Holstein-Hubbard system
  (\ref{hhm}) at $U=3$, $t=1$, $\omega_0=1$, $g^2=2$ and periodic BCs,
  has been chosen to show the
  deficiencies of the parallelization approach. We represent each boson site 
  with six pseudosites~\cite{jec98} corresponding to a maximum of 64 phonons 
  per boson site. Thus, the effective number of DMRG sites is 56\@. To achieve 
  convergence $m=900$ has to be used\@.
\end{enumerate}
Although this study deals mainly with scalability, we nevertheless
specify one-CPU performance numbers for all systems under
investigation in order to set the scale (Table~\ref{onecpu})\@.
\begin{table}\renewcommand{\arraystretch}{1.5}
\caption{\label{onecpu}\bunt One-CPU performance in GFlop/s and 
efficiency in terms of fraction of peak performance for
all systems studied (benchmark case \ref{case1})\@. Proprietary,
vendor-supplied BLAS and LAPACK implementations were used in
all cases.}
\begin{center}
\begin{tabular}{l|c|c|c}\hline
\multicolumn{1}{p{1.8cm}|}{\textbf{System}} &
\multicolumn{1}{p{2.2cm}|}{\raggedright\textbf{Peak Perf. [GFlop/s]}} &
\multicolumn{1}{p{2.6cm}|}{\raggedright\textbf{DMRG Perf. [GFlop/s]}} & 
\multicolumn{1}{p{1.8cm}}{\raggedright\textbf{Fraction of Peak}} \\\hline 
IBM p690/Power4 (1.3\,GHz) & 5.2 & \textbf{2.78} & 0.53 \\
HP rx5670/Itanium2 (1\,GHz) & 4.0 & \textbf{2.25} & 0.56 \\
Intel Xeon DP (2.4\,GHz) & 4.8 & \textbf{2.08} & 0.43 \\
SunFire 3800 (900\,MHz) & 1.8 & \textbf{0.92} & 0.51 \\
SGI Origin 3400 (500\,MHz) & 1.0 & \textbf{0.78} & 0.78 \\\hline
\end{tabular}
\end{center}
\end{table}
Although it is clear that performance is always dominated by the
Davidson diagonalization, the quality of the C++ compiler and the
DGEMM implementation have some influence, the latter especially due to
the abundance of small and non-square matrices. Because of a
sophisticated, object-oriented data housekeeping structure in the
code, proper inlining and optimization is essential as well.  A
comparison with peak performance for every system (last column in
Table~\ref{onecpu}) shows deficiencies in those respects quite prominently.

For parallel performance studies there are essentially two metrics
that can be considered: Speedup $S(N)$ and parallel efficiency
$\varepsilon(N)$\@. If $P(N)$ is the performance of the benchmark
on $N$ processors, then
\bq
S(N)={P(N)\over P(1)}\qquad\mbox{and}\qquad\varepsilon(N)={S(N)\over N}\eos
\eq
In the following we will present data for one or the other metric 
as appropriate.

An important limitation to parallel efficiency and speedup is imposed
by a theoretical limit called \emph{Amdahl's Law}.  In a simple model
one can split a single-threaded application into a serial
(non-parallelizable) fraction $s$ and a perfectly parallelizable
fraction $p=1-s$\@. The speedup with $N$ CPUs is then calculated
as
\bq
S_{\mathrm A}(N) = {s+p\over s+{p\over N}} = {1\over s+{1-s\over N}}\cma
\eq
with
\bq
\lim_{N\to\infty}S_\mathrm{A}(N)={1\over s}\eos
\eq
In our case the serial fraction is strongly influenced by the quality
of the C++ compiler, which has thus a large impact on scalability.  As
already mentioned in section \ref{omppar}, the typical fraction of
85\,\% of the total computing time for the sparse MVM (leading to
$p=0.85$ in the Amdahl model) leads to the expectation that speedups
between 6 and 7 are achievable when parallelization overhead is
negligible.

\subsection{Parallel DGEMM}

\begin{figure}
\centerline{\includegraphics*[width=0.85\textwidth]{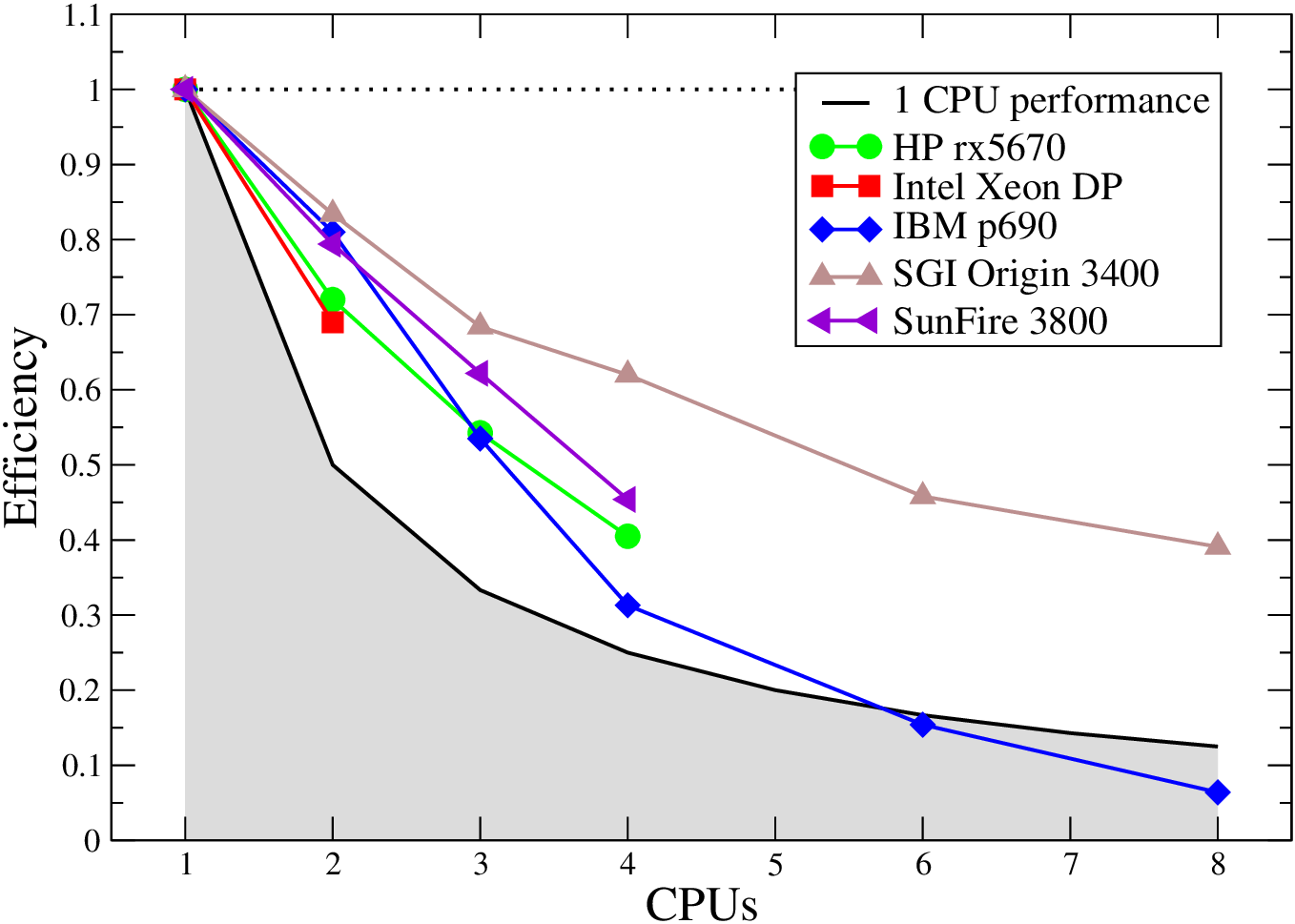}}
\caption{\label{dgemmeff}\bunt Parallel efficiency on different SMP systems 
  (whole program) with DGEMM parallelization, benchmark case \ref{case1}\@.
  The grey region marks 
  the limit where no speedup is gained compared to the 1-CPU run.}
\end{figure}

Using parallel DGEMM is as easy as relinking with the appropriate
library on all systems, and is available everywhere. Parallel
efficiency was measured on a variety of architectures (see
Fig.~\ref{dgemmeff})\@.
As can be seen from the parallel efficiency data, this is actually a
very poor method for parallelization. Scalability depends heavily on
the quality of the implementation of parallel DGEMM, as well as more
obscure features like hardware barriers and associated loss. Compared
to other systems, the SGI Origin still does quite well, which can at
least partly be attributed to the high-quality C++ compiler.

Fig.~\ref{dgemmeff} also shows the limit where parallelization
becomes entirely useless (grey zone), i.~e.\ where $N$-CPU
performance drops below the 1-CPU case.

\subsection{OpenMP Parallelization}

The OpenMP variant of the program unfortunately runs only with
SGI and IBM systems. Intel and Sun compilers have deficiencies that
either prevent the code from compiling or generate nonfunctional
programs.

\begin{figure}
\centerline{\includegraphics*[width=0.85\textwidth]{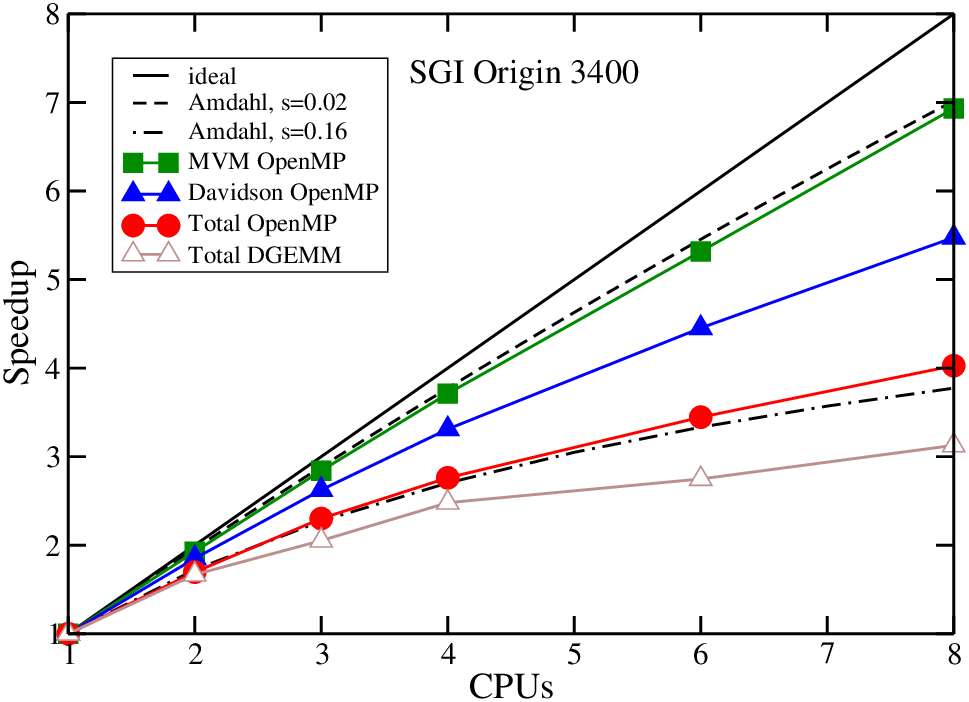}}
\caption{\label{sgiscal}\bunt OpenMP scaling on SGI Origin 3400, benchmark
  case \ref{case1}\@. Scaling
  of whole program, Davidson algorithm and MVM are shown separately,
  as well as Amdahl scaling for serial fractions $s=0.05$ and
  $s=0.16$\@.}
\end{figure}

Fig.~\ref{sgiscal} shows the results of a scaling run with up to 8
CPUs on an Origin 3400 system, where scaling is broken down to
different abstraction levels (MVM, Davidson, whole program)\@. While
the ``whole program'' scaling is what the end user is finally
interested in, it is quite clear that some significant optimization
potential is still hidden between Davidson diagonalization and sparse
MVM\@.
Amdahl scaling for two different serial fractions ($s=0.02$ and
$0.16$) is also shown.  Although the Amdahl performance model is
admittedly too simplistic for this code, it nevertheless gives a rough
impression about what has been achieved. Obviously, the MVM
parallelization is very efficient with only a minor serial fraction.

\begin{figure}
\centerline{\includegraphics*[width=0.85\textwidth]{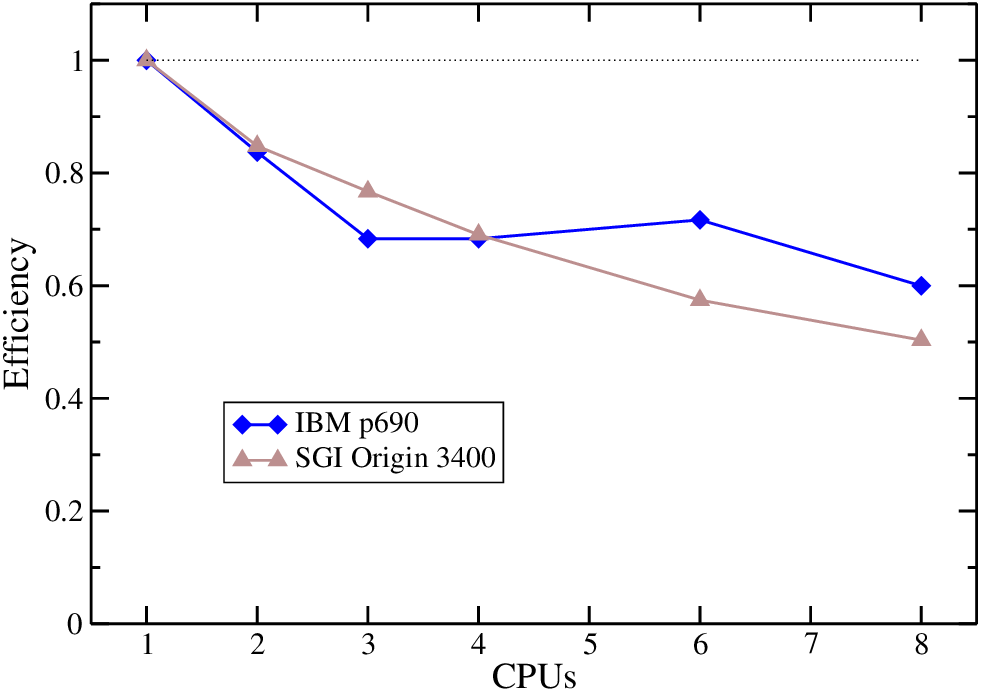}}
\caption{\label{ompeff}\bunt OpenMP parallel efficiency on IBM p690
  and SGI Origin SMP systems (whole program), benchmark case \ref{case1}\@.
  The data for the IBM system was taken on a loaded system.}
\end{figure}
\begin{figure}
\centerline{\includegraphics*[width=0.85\textwidth]{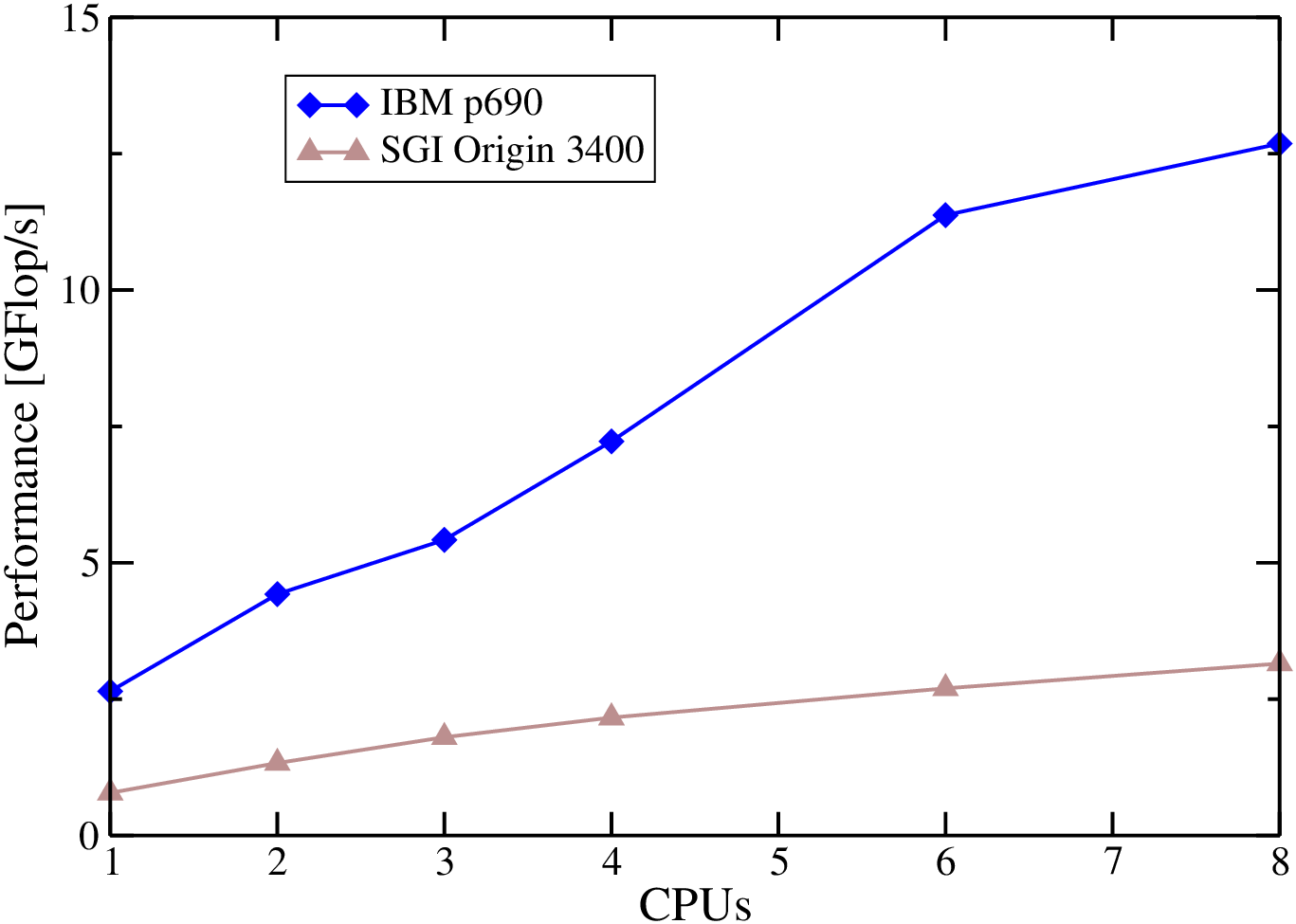}}
\caption{\label{omptot}\bunt OpenMP absolute performance in GFlop/s
  on IBM p690 and SGI Origin SMP systems (whole program), benchmark 
  case \ref{case1}\@.}
\end{figure}

Fig.~\ref{ompeff} displays the parallel efficiency of the code 
on IBM p690 and SGI Origin 3400 systems. In contrast to the DGEMM
parallelization case, SGI does not have an advantage here.
Although the two systems are practically on par with respect to
scalability, a direct comparison of performance in GFlop/s
shows clearly what the favourable architecture for DMRG today
should be (Fig.~\ref{omptot})\@.

\begin{figure}
\centerline{\includegraphics*[width=0.85\textwidth]{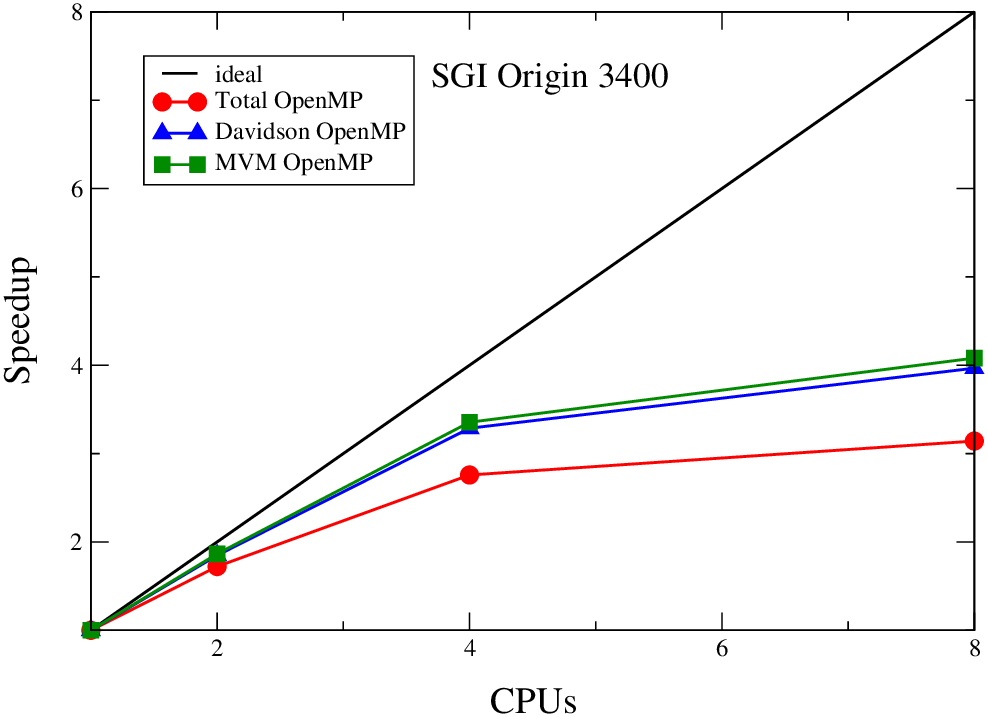}}
\caption{\label{ompeffhhm}\bunt OpenMP scaling for benchmark 
case \ref{case2}\@.}
\end{figure}

As the Davidson procedure itself is very well parallelizable, we
expect that some performance boost is still in reach. Other aspects of
the implementation that become more prominent with other physical
setups also bear some optimization potential. An example for this is
the Holstein-Hubbard model (benchmark case \ref{case2}) for which 
the broken-down parallel profiling data is shown in Fig.~\ref{ompeffhhm}\@.
Here we see that the mediocre overall speedup is actually caused by
the sparse MVM itself. Profiling reveals that a significant amount of
time is spent in acquiring locks for the parts of the result vector.
Reordering the loop iterations may help here and is being investigated.

\section{Application: Peierls-Insulator Mott-Insulator Transition
in 1D}\label{phys}

\begin{figure}[tbh!]
\centerline{\includegraphics*[width=0.85\textwidth]{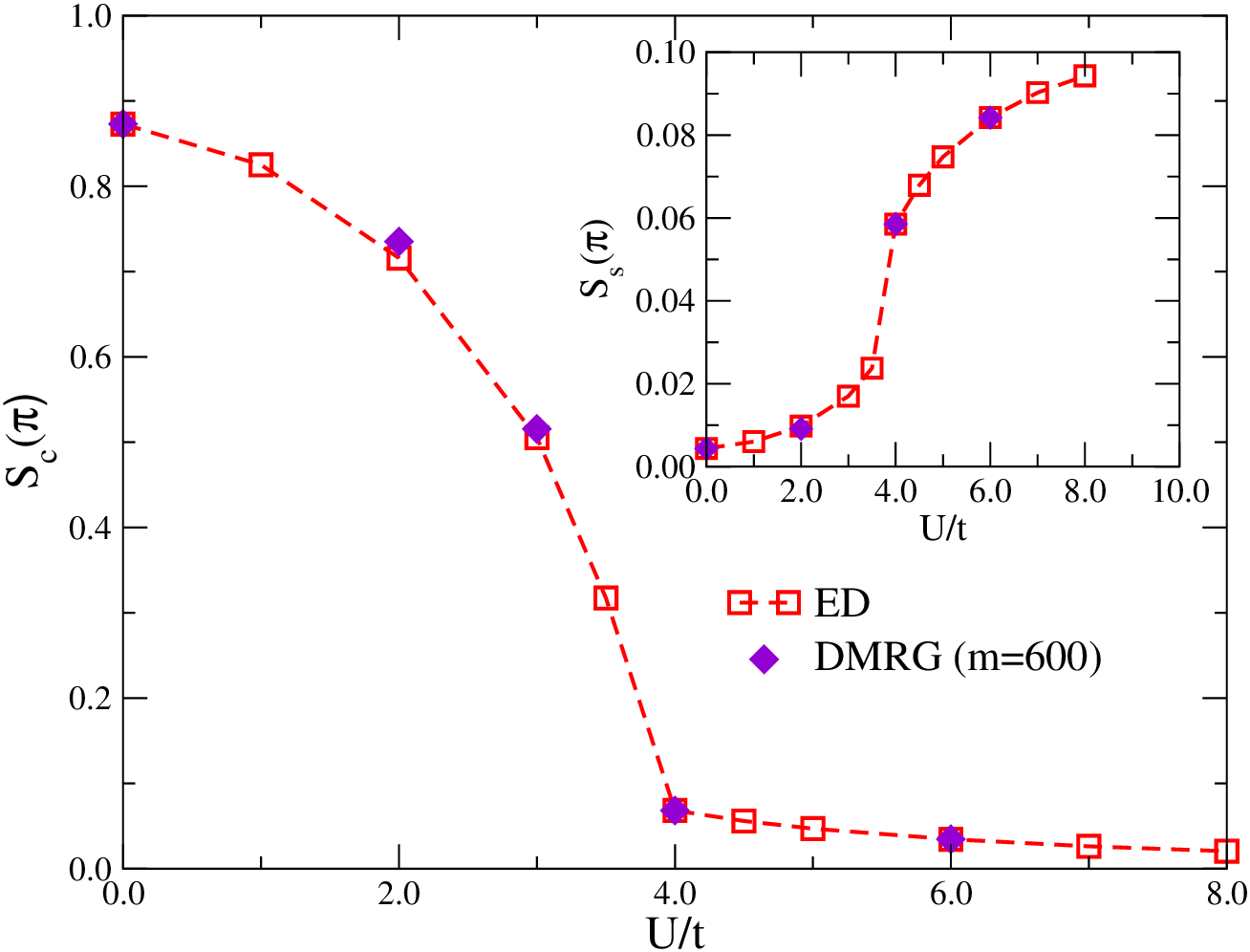}}
\caption{\label{f1col}\bunt Spin and charge 
  structure factors at $q=\pi$ in the half-filled one-dimensional 8-site HHM
  (\ref{hhm}) with periodic BCs for different $U$ at $t=1$, $\omega_0=1$ and
  $g^2=2$\@. Squares denote ED results, diamonds show DMRG calculations
  with $m=600$ and six pseudosites.}
\end{figure}

\begin{figure}[tbh!]
\centerline{\includegraphics*[width=0.85\textwidth]{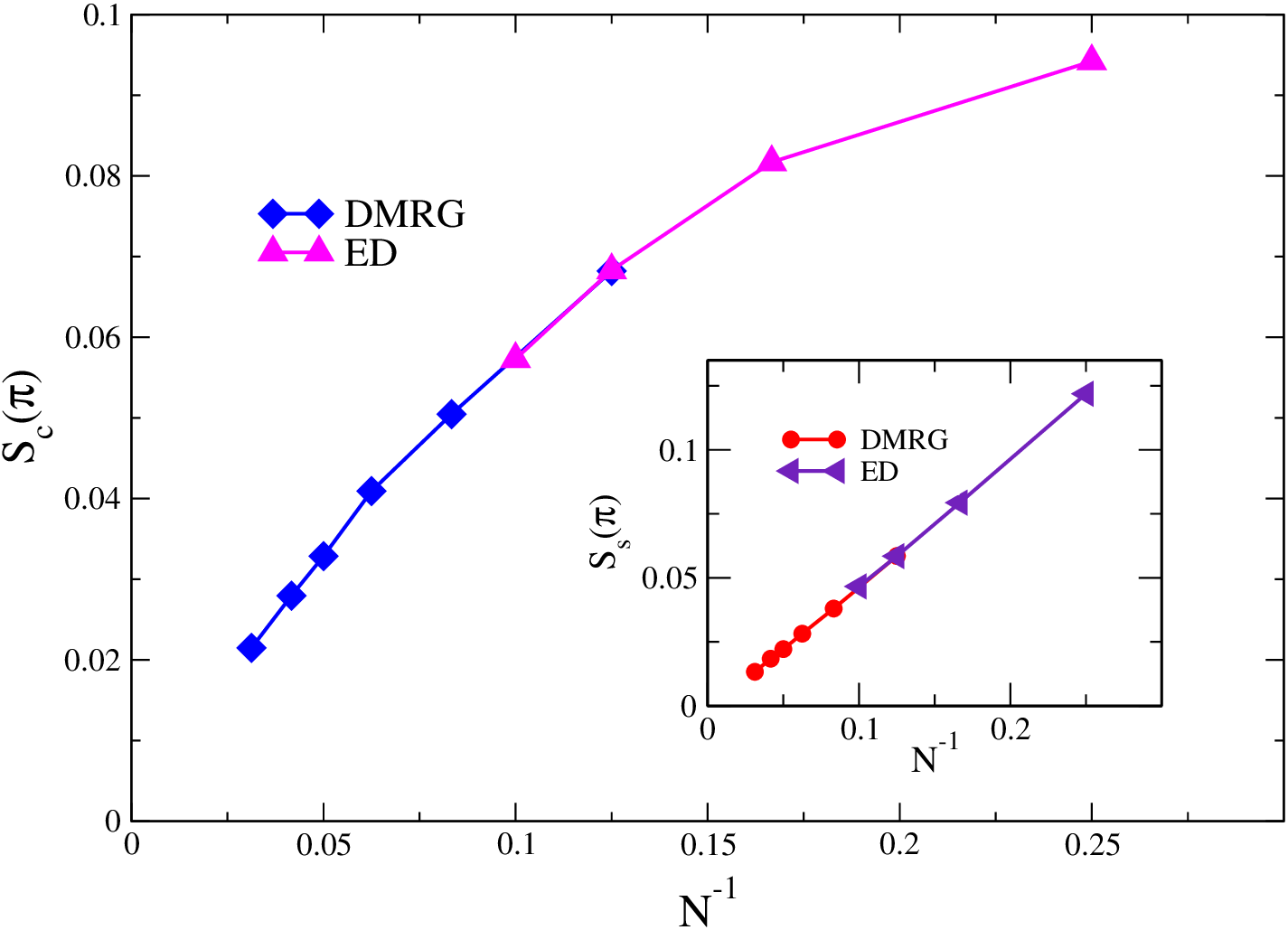}}
\caption{\label{chsp}\bunt Finite-size scaling study of spin and charge 
  structure factors at $q=\pi$ in the half-filled one-dimensional HHM
  (\ref{hhm}) with periodic BCs at $U=4$, $t=1$, $\omega_0=1$ and
  $g^2=2$ with five boson pseudosites, $m=1000$ and lattice sizes of up to
  32\@. For reference, available ED calculations are shown as well.}
\end{figure}

In quasi-one-dimensional (1D) materials there is a strong competition
between electron-electron and electron-phonon interactions, which tend
to localize the charge carriers by establishing commensurate
spin-density-wave and charge-density-wave ground states, respectively.
At half-filling, in particular, Peierls (PI) or Mott (MI) insulating
phases are favored over the metallic state.  A heavily debated issue
concerns the nature of the quantum phase transitions between the
different insulating phases (for more details see~\cite{FKSW03} and
references therein)\@.  The Holstein-Hubbard model is perhaps the most
simple model to address this problem because it shows a PI-MI
transition with increasing $U$ above a threshold electron-phonon
coupling (a critical electron-phonon coupling is required in order to
establish the PI state at nonzero phonon frequency).  For finite
periodic chains it has been verified that the transition results from
a ground state level crossing with a change in the ground state
site-parity eigenvalue. As can be seen from Fig.~\ref{f1col}, 
the staggered charge- and spin-structure factors,
\begin{eqnarray}
S_{c}(\pi) &=& \frac{1}{N^2}\sum_{i,j\atop\sigma\sigma'} (-1)^{|i-j|}\langle
(n_{i\sigma}-\frac{1}{2} )(n_{j\sigma'}-\frac{1}{2})\rangle\cma\\
S_{s}(\pi) &=& \frac{1}{N^2}\sum_{i,j} (-1)^{|i-j|}\langle S_i^zS_j^z\rangle\cma
\quad S_i^z={1\over 2}(n_{i\uparrow}-n_{i\downarrow})\cma
\label{scs}
\end{eqnarray}
are strongly suppressed approaching the quantum critical point from
below and above, respectively. However, both $S_{c}(\pi)$ and $S_{s}(\pi)$
remain finite at the transition point for the small 8-site system
we were able to study by means of ED techniques in previous work. 
Fig.~\ref{f1col} shows that good agreement between ED and DMRG
is achievable for this case.
Using the parallelized DMRG code for the Holstein-Hubbard model
we are now in the position to calcluate spin and charge structure
factors for a sequence of systems with up to 32 sites. 
The results presented in Fig.~\ref{chsp} can be used to perform 
a reliable finite-size extrapolation: At the quantum citical point
$S_{c}(\pi)$ and $S_{s}(\pi)$  vanish in the thermodynamic 
limit $N\to\infty$. Simultaneously the optical excitation gap closes.

A comparison of the required resources for this problem shows already in
the 8-site case quite clearly the superior capabilities of the
DMRG method for this kind of problem (Table~\ref{rescomp})\@.
\begin{table}[tbp]\renewcommand{\arraystretch}{1.5}
\setlength{\tabcolsep}{1mm}
\caption{\label{rescomp}\bunt Comparison of computational resources
for the calculation of spin and charge structure factors in the Holstein-Hubbard
Model}
\centerline{%
\begin{tabular}{l|c|c|c}\hline
\textbf{Method} & \textbf{\# of CPUs} & \textbf{Walltime} & \textbf{Memory} \\\hline
ED (8 sites, matrix dim.\ $\sim10^{10}$) & 1024 (Hitachi SR8000) & 
$\sim12$\,hrs & 600\,GB \\
DMRG (8 sites, $m=600$) & 1 (SGI Origin) & $\sim18$\,hrs & 2\,GB \\
DMRG (24 sites, $m=1000$) & 4 (SGI Origin) & $\sim72$\,hrs & 10\,GB\\\hline
\end{tabular}%
}
\end{table}
For the 32-site lattice (leftmost data point in Fig.~\ref{chsp}) with
five pseudosites (32 phonons per boson site), the overall number of sites is
192\@. Such a system would be absolutely unmanageable with ED methods.

\section{Conclusions and Outlook}\label{conc}

We have presented two methods for parallelization of a DMRG code on
shared-memory systems: parallel DGEMM and OpenMP parallelization 
on the Davidson MVM level. The deficiencies of parallel DGEMM 
are quite clear, but it is still the only alternative when one 
has to stick to compilers that do
not support OpenMP directives (correctly)\@. OpenMP does much better,
and there is some significant parallelization potential still hidden in
the code outside the MVM subroutine that must be exploited. 
We expect that the parallel code will 
scale well up to sixteen CPUs without any changes in the DMRG algorithm.
A radically different parallelization approach or a new DMRG algorithm 
would be necessary to obtain reasonable scalability on hundreds of processors 
in massively parallel computers.
However, the current SMP implementation already allows us to investigate much 
larger systems than with ED or with sequential DMRG.  

\section*{Acknowledgements}

We thank the RRZE, the RZG (Computing Center Garching), the URZ
(Universit\"atsrechenzentrum Dresden) and the HLRS (High Performance
Computing Center Stuttgart) for providing computational resources.
The RRZN (Regionales Rechenzentrum Niedersachsen) and the ZIB
(Zuse-Institut Berlin) have granted resources on their HLRN
supercomputer complex. This work was partially supported by the
Bavarian Competence Network for High Performance Computing
(KONWIHR)\@.

\end{document}